\documentclass{Interspeech}

\interspeechcameraready

\title{Attractor-Based Speech Separation of Multiple Utterances \\ by Unknown Number of Speakers}

\author[affiliation={1}]{Yuzhu}{Wang}
\author[affiliation={1}]{Archontis}{Politis}
\author[affiliation={2}]{Konstantinos}{Drossos}
\author[affiliation={1}]{Tuomas}{Virtanen}

\affiliation{Signal Processing Research Center}{Tampere University}{Finland}
\affiliation{}{Nokia Technologies}{Finland}
\email{yuzhu.wang@tuni.fi, archontis.politis@tuni.fi, konstantinos.drosos@nokia.com, tuomas.virtanen@tuni.fi}

\usepackage{comment}
\usepackage{physics}
\begin{document}

\maketitle
% the abstract here must exactly match the abstract entered into the paper submission system
\begin{abstract}
% 1000 characters. ASCII characters only. No citations.
This paper addresses the problem of single-channel speech separation, where the number of speakers is unknown, and each speaker may speak multiple utterances. 
We propose a speech separation model that simultaneously performs separation, dynamically estimates the number of speakers, and detects individual speaker activities by integrating an attractor module.
The proposed system outperforms existing methods by introducing an attractor-based architecture that effectively combines local and global temporal modeling for multi-utterance scenarios.
To evaluate the method in reverberant and noisy conditions, a multi-speaker multi-utterance dataset was synthesized by combining Librispeech speech signals with WHAM! noise signals.
The results demonstrate that the proposed system accurately estimates the number of sources. The system effectively detects source activities and separates the corresponding utterances into correct outputs in both known and unknown source count scenarios. 
\end{abstract}

\vspace{1mm}
\section{Introduction}
Speech separation aims to isolate individual speech signals from mixtures containing multiple speakers. This task is particularly challenging in real-world environments due to noise, reverberation, and the highly time-varying nature of source activities, especially in single-channel scenarios~\cite{vincent2018audio, purwins2019deep, cobos2022overview, wang2018supervised}. 

Most existing research assumes a known and fixed number of speakers with each contributing a single utterance~\cite{luo2019conv, luo2020dual, Yu2017permutation, kolbaek2017multitalker, Subakan2021, Chen2020DPTnet, Wang2023TFGridNet, Wang2023TFGridNet2}. However, this assumption does not match with real-world scenarios, where the number of speakers is often unknown and varies over time, with each speaker potentially contributing multiple utterances that may overlap in variable ways.
To handle unknown source numbers, one direct approach is recursive separation~\cite{Takahashi2019, Kinoshita2018}.
In contrast, deep clustering methods~\cite{Hershey2016, wang2019low} create high-dimensional embeddings for each time-frequency bin, which are then clustered to group together the embeddings corresponding to the same speaker, thereby avoiding the dependency on the prior number of speakers.
A similar approach is applied in speaker clustering~\cite{Zeghidour2020}.
Deep attractor networks~\cite{Chen2017deep, Luo2018speaker} try to create a representation vector for each sound source in feature space as a reference point.
Following the success of attractor networks, systems that combine attractors and separators have gradually emerged as the mainstream solution.
RNN-based attractors were explored in joint speaker diarization and speech separation systems~\cite{Maiti2023EENDSS, Chetupalli2022}.
More recently, a transformer attractor was proposed in~\cite{lee2024boosting}, which demonstrated excellent performance.

The aforementioned studies focus only on single-utterance scenarios.
For situations with an unknown number of speakers and multiple utterances, several methods have shown potential effectiveness, including recurrent attention networks~\cite{zhang2022continuous}, spatial filtering~\cite{zhang2022all}, and dual-path model consisting of LSTM and transformers~\cite{raj2022continuous}.
However, most of these approaches have concentrated on speaker counting and long-sequence modeling, overlooking source activity detection.
Moreover, they primarily target speech recognition, with limited evaluations of separation performance and the impact of noise and reverberation.

\begin{figure}
\vspace{-1mm}
\centering
\includegraphics[width=0.40\textwidth]{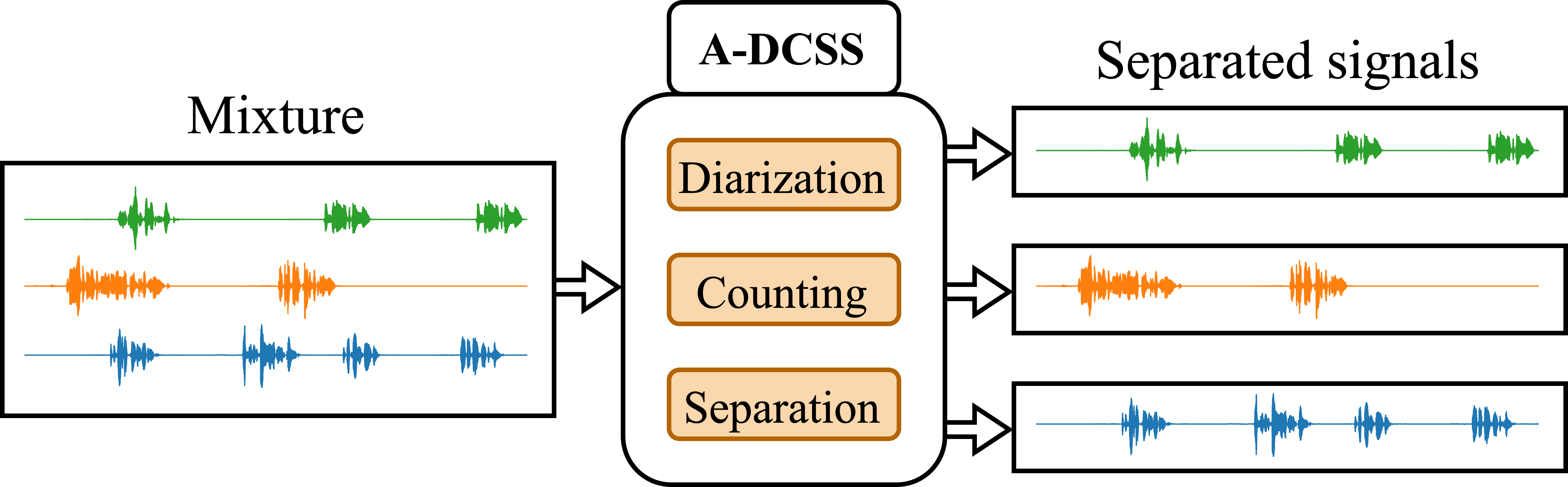}
\vspace{-2mm} 
\caption{The illustration of the A-DCSS in multi-speaker with multi-utterance scenarios.}
\label{fig-problem}
\vspace{-6mm}
\end{figure}
Our work aims to address the novel problem of separating an unknown number of speakers, each with multiple utterances, a scenario that has not been evaluated in previous research, as shown in Fig.~\ref{fig-problem}. 
Related research exists in continuous speech separation (CSS)~\cite{chen2020continuous, li2021dual}, which focuses on producing output streams where speech signals do not overlap, typically assuming a known number of sources. In contrast, our work emphasizes separation of unknown number of speakers while ensuring all utterances from the same speaker are consistently assigned to a single output channel, rather than merely de-overlapping the signals.
We propose an attractor-based joint diarization, counting, and separation system (A-DCSS) that leverages multi-task learning to simultaneously minimize errors across these tasks.
Our system makes four major contributions: 
(1) We are the first to test an attractor design for multi-utterance separation, validating its effectiveness in source modeling and potential source identification. 
(2) We integrate RNN attractors with feature-wise linear modulation (FiLM)~\cite{Perez2018film} and develop an improved multi-path module with transformer-LSTM blocks~\cite{Chen2020DPTnet}. 
% and develop an improved multi-path module with transformer-LSTM blocks.
(3) Our ablation studies demonstrate that source activity detection significantly improves separation performance. 
(4) We retrain and evaluate existing methods~\cite{Takahashi2019, Maiti2023EENDSS, Chetupalli2022, lee2024boosting} on the synthesized dataset, demonstrating their adaptability to multi-utterance scenarios.
Our experimental results show that A-DCSS outperforms baseline systems in both fixed and varying speaker scenarios under anechoic, noisy, and reverberant conditions.
\vspace{-2mm}
\section{Signal Model}
\label{sec-signal-model}
%We introduce a joint signal model of speaker diarization, counting, and separation. 
The observed signal $\mathbf{y}$ of length $T^{*}$ can be represented as
\vspace{-2mm}
\begin{equation}
\label{eq:observed_signal}
\mathbf{y} = \sum_{c=1}^{C} \mathbf{x}_{c} + \mathbf{n}.
% \vspace{-1mm}
\end{equation}
Here, the observed signal $\mathbf{y}$ consists of speech signals $\mathbf{x}_{c}$ from $C$ sources and noise $\mathbf{n}$.
The signal $\mathbf{x}_{c}$ from the $c$-th speaker, consists of multiple utterances and potential pauses between them.
The speaker counting task involves estimating the number of speakers $C$.
We assume that by extracting features from the time-domain signals, we can predict binary speaker activities $\hat{\mathbf{P}}$ of size $C \times T$, where $T$ denotes the number of time segments in which the activity is predicted.
In realistic scenarios
the number of speakers $C$ is unknown, and the speech signal $\mathbf{x}_{c}$ consists of multiple utterances. 
Our objective is to estimate $\mathbf{x}_{c}$ by separating all utterances belonging to the same speaker into a single output.
This process leverages the estimated speaker count $\hat{C}$ and speaker activity $\hat{\mathbf{P}}$, resulting in $\hat{C}$ separated speech signals, as illustrated in Fig.~\ref{fig-problem}.
\begin{figure}
\centering
\includegraphics[width=0.42\textwidth]{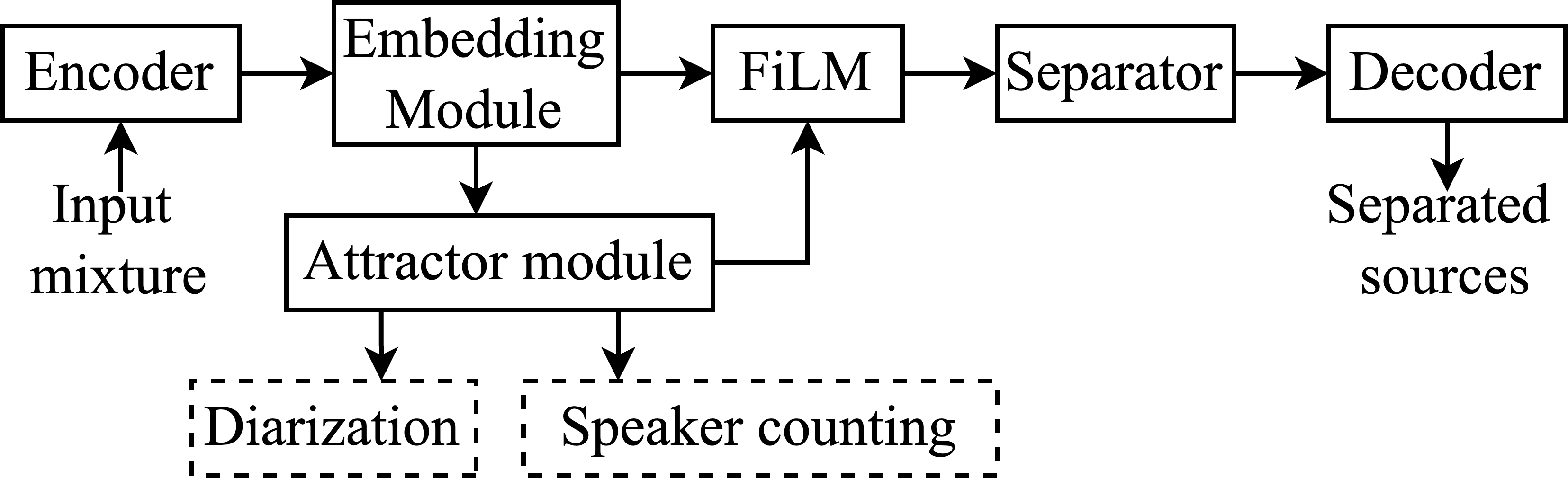}
\vspace{-1mm}
\caption{A-DCSS architecture.}
\label{fig-system}
\vspace{-6mm}
\end{figure}
% \vspace{-1ex}
\section{Attractor-based joint diarization, counting, and separation system (A-DCSS) }
\label{sec-proposed-system}
% \subsection{System Architecture}
The proposed joint separation system consists of three core processes: feature embedding, attractor generation, and speech separation. The system architecture is illustrated in Fig.~\ref{fig-system}. 
An encoder and a decoder are used to transform the signal between the time domain and the time-feature domain.
The embedding module maps the time-feature representations into a latent embedding space through a dual-path transformer-based network.
The attractors are generated by learning source-level information from the embeddings.
As shown in Fig.~\ref{fig-system}, the speaker diarization results and the prediction of the attractor existence probability can be obtained by the attractors, respectively.
Then, the generated attractors are used as conditioning input, modulating the mixture features into speaker-specific features through a FiLM operation.
Finally, we employ a triple-path separator to achieve speech separation. 

%\vspace{-1ex}
%\subsection{Encoder and decoder}
\noindent\textbf{Encoder and decoder}: The encoder applies a one-dimensional convolutional layer with rectified linear unit (ReLU) activation to transform the input 
%waveform 
$\mathbf{y} \in \mathbb{R}^{T^*}$ into a time-feature representation $Y \in \mathbb{R}^{T \times F}$, 
$F$ being the feature dimension.
The kernel size is $L$ samples and the stride is $L/2$ samples.
The decoder reconstructs each separated signal using a transposed convolutional layer with the same hyperparameters as the encoder.

\begin{figure}
\centering
\includegraphics[width=0.42\textwidth]{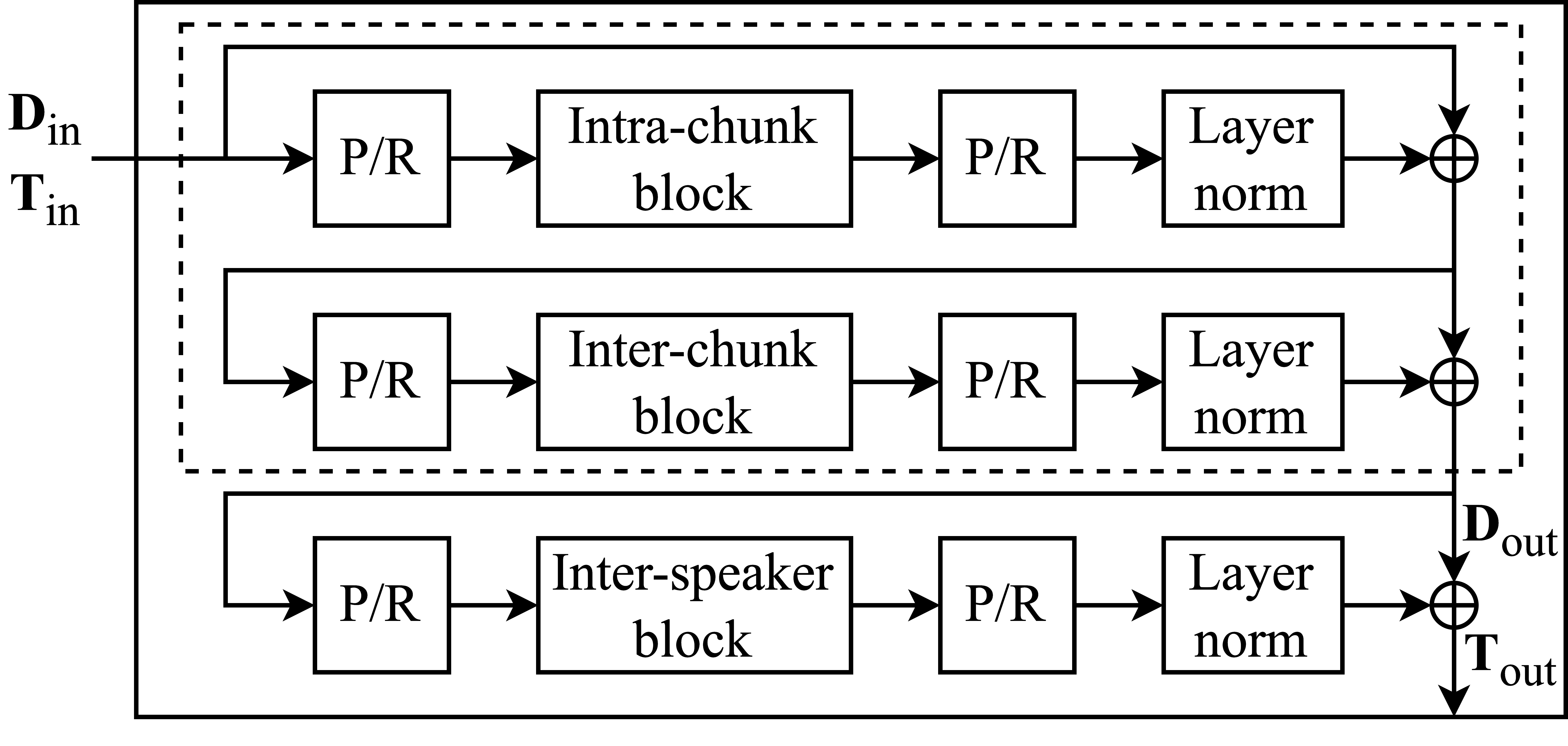}
\vspace{-.5em}
\caption{Dual-path (dashed line) and triple-path modules. $\mathbf{D}_{\text{out}}$ and $\mathbf{T}_{\text{out}}$ are their outputs with inputs $\mathbf{D}_{\text{in}}$ and $\mathbf{T}_{\text{in}}$, respectively. $\text{P}$ and $\text{R}$ represent the permute and reshape operations.}
\label{fig-dual-and-triple-path}
% \vspace{-.5cm}
\vspace{-6mm}
\end{figure}

%\vspace{-1ex}
%\subsection{Feature Embedding}
\noindent\textbf{Feature Embedding}: The feature embedding module of A-DCSS uses a dual-path design, as shown in Fig.~\ref{fig-dual-and-triple-path}.
A linear layer is used to map the encoded features from dimension $F$ to $D$. Following the Sepformer~\cite{Subakan2021}, we employ a chunking operation that segments the time dimension $T$ into $S$ chunks of equal length $K$, with a hop size of $K/2$, resulting in the input tensor $\mathbf{D}_{\text{in}} \in\mathbb{R}^{K \times S \times D}$. This dual-path structure allows the module to model both local dependencies within chunks through intra-chunk processing ($S \times K \times D$) and global dependencies across chunks through inter-chunk processing ($K \times S \times D$).
In Fig.~\ref{fig-dual-and-triple-path}, $\text{P}/\text{R}$ represents the permute/reshape operations. The intra-chunk and inter-chunk blocks follow the design in~\cite{Subakan2021}, consisting of a multi-head attention layer followed by a feed-forward layer, with layer normalization before each layer.

\begin{figure}
\centering
\includegraphics[width=0.42\textwidth]{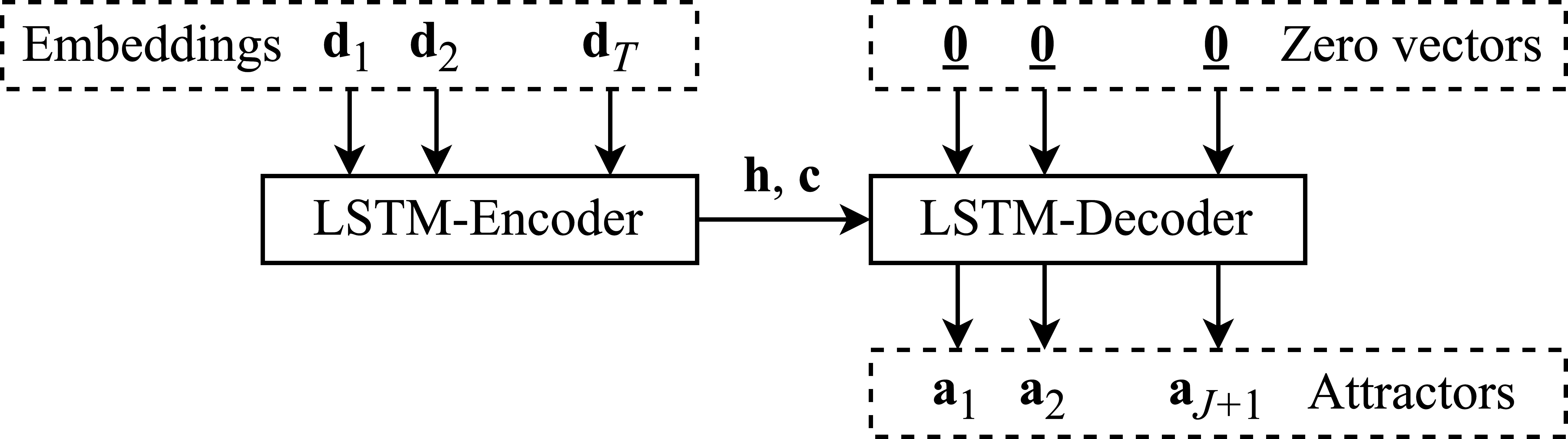}
\vspace{-2mm}
\caption{Attractor module.}
\label{fig-attractor}
\vspace{-6mm}
\end{figure}

%\vspace{-1ex}
%\subsection{Attractor module}
\noindent\textbf{Attractor module}: The attractor module aims to learn source-level representations that characterize individual speakers in the mixture. Unlike the attractors in~\cite{Chen2017deep, Luo2018speaker} obtained through ideal mask and k-means clustering, 
our approach adopts a BLSTM-based encoder-decoder architecture for attractor generation as shown in Fig.~\ref{fig-attractor}, following~\cite{horiguchi2020end}. 
The output $\mathbf{D}_{\text{out}} \in\mathbb{R}^{K \times S \times D}$ from the dual-path module is processed by an overlap-add operation, reconstructing it into a sequence $\{\mathbf{d}_t\}_{t=1}^{T} \in\mathbb{R}^{T \times D}$.
The LSTM-encoder encodes the obtained sequence to generate hidden state $\mathbf{h}\in\mathbb{R}^D$ and cell states $\mathbf{c}\in\mathbb{R}^D$, which are subsequently used by the LSTM-decoder to generate $J+1$ attractors $\{\mathbf{a}_j\}_{j=1}^{J+1}$, where $\mathbf{a}_j \in\mathbb{R}^{1 \times D}$. 
During the decoding step, a sequence of $D$-dimensional zero vectors $\{\mathbf{h}_{\underline{\mathbf{0}}}\} \in\mathbb{R}^{(J+1) \times D}$ is used as input. 
The first $J$ attractors $\{\mathbf{a}_j\}_{j=1}^{J} \in\mathbb{R}^{J \times D}$ represent the identified $J$ sources, while the final attractor $\mathbf{a}_{J+1}$ is responsible for determining the non-existence of the speaker.
As shown in Fig.~\ref{fig-system}, the attractor module produces three outputs: 
(1) The generated attractors $\{\mathbf{a}_j\}_{j=1}^{J+1} \in\mathbb{R}^{(J+1) \times D}$ are passed through a linear layer with a sigmoid activation function to get the prediction of the attractor existence probability $\hat{\mathbf{q}} \in\mathbb{R}^{J+1}$.
Binary prediction values are obtained by comparing each element of $\hat{\mathbf{q}}$ with the predefined threshold $\tau_{\text{exist}}$.
The estimated number of speakers $\hat{C}$ corresponds to the number of existing attractors;
(2) The inner product between the first $J$ attractors $\{\mathbf{a}_j\}_{j=1}^{J} \in\mathbb{R}^{J \times D}$ and the input embeddings $\{\mathbf{d}_t\}_{t=1}^{T} \in\mathbb{R}^{T \times D}$ are passed through a linear layer with a sigmoid activation function to predict the probability of activity of each source at each time step, which is represented using matrix $\hat{\mathbf{P}} \in\mathbb{R}^{J \times T}$. This process follows the attractor-based diarization method in~\cite{horiguchi2020end}.
Binary activity predictions are obtained by comparing each element of $\hat{\mathbf{P}}$ with the predefined threshold $\tau_{\text{diar}}$;
(3) The generated first $J$ attractors $\{\mathbf{a}_j\}_{j=1}^{J} \in\mathbb{R}^{J \times D}$ are fused using FiLM conditioning~\cite{Perez2018film} with the dual-path module output $\mathbf{D}_{\text{out}} \in\mathbb{R}^{K \times S \times D}$ to get the input of separator $\mathbf{T}_{\text{in}} \in\mathbb{R}^{J \times K \times S \times D}$.
During training, $J$ is set to the true number of speakers $C$. During inference, we define a maximum possible number of sources $J_{\text{max}}$ and generate $J_{\text{max}}+1$ attractors. Then, an estimate $\hat{C}$ is obtained based on the predicted probability of attractor existence, and the first $\hat{C}$ attractors are passed to the speaker diarization and separation branches.

%\vspace{-1ex}
%\subsection{Triple-path Separator}
\noindent\textbf{Triple-path Separator}: 
The triple-path module extends the dual-path module by incorporating an additional inter-speaker block, following the design in~\cite{lee2024boosting}, as shown in Fig.~\ref{fig-dual-and-triple-path}.
After FiLM modulation, the feature tensor $\mathbf{T}_{\text{in}} \in\mathbb{R}^{J \times K \times S \times D}$ incorporates an additional dimension $J$ corresponding to the first $J$ attractors. 
The inter-speaker block processes information along the $J$ dimension, treating the $K$ and $S$ as a batch dimension.
Each block of the triple-path module employs an enhanced transformer-LSTM block instead of the simple transformer block used in the dual-path module.
This enhanced block, which has demonstrated superior performance for long speech sequence modeling~\cite{Chen2020DPTnet}, comprises a multi-head attention layer, an LSTM layer, and a linear layer.
% Residual connections are applied, with layer normalization following each sub-layer.
Furthermore, the three-path module is stacked 
$N_{\text{triple}}$ times, which enhances the capacity to capture complex patterns.
The output of the triple-path module $\mathbf{T}_{\text{out}} \in\mathbb{R}^{J \times K \times S \times D}$ is first processed by overlap-add to obtain $\mathbf{T'}_{\text{out}} \in\mathbb{R}^{J \times T \times D}$. Then, a linear output layer transforms the feature dimension from \( D \) to \( F \), and the $J$ separated time-domain signals are obtained after decoding.
% \vspace{-2mm}
% \subsection{Loss Function}
\noindent\textbf{Loss Function}: A-DCSS is trained using a joint loss function,
$\mathcal{L}_{\text{joint}} = \lambda_{\text{s}} \mathcal{L}_{\text{SI-SDR}} + \lambda_{\text{d}} \mathcal{L}_{\text{diar}} + \lambda_{\text{e}} \mathcal{L}_{\text{exist}}$.
$\lambda_{\text{s}}$, $\lambda_{\text{d}}$, and $\lambda_{\text{d}}$ denote the weights assigned to each loss item.
Here, $\mathcal{L}_{\text{SI-SDR}}$ is the SI-SDR loss that measures the reconstruction error between the separated signal $\hat{\mathbf{x}}$ and reference signal $\mathbf{x}$.
$\mathcal{L}_{\text{diar}}$ represents the binary prediction loss for speaker activity.
The permutation invariant training (PIT) method~\cite{Yu2017permutation, kolbaek2017multitalker} is used to calculate the diarization loss as 
\vspace{-1mm}
\begin{equation} \label{eq:loss_diar}
\mathcal{L}_{\text{diar}}=\underset{\text{PIT}}{\min} \underset{t}{\sum}~\text{BCE}(\mathbf{p}_{t},\hat{\mathbf{p}}_{t}),
% \vspace{-1mm}
\end{equation}
where $\mathbf{p}_{t}$ is a vector of reference labels at time $t$.
$\text{BCE}(\cdot, \cdot)$ is the binary cross entropy loss.
$\mathcal{L}_{\text{exist}}$ is the attractor existence loss, $\mathcal{L}_{\text{exist}}=\text{BCE}(\mathbf{q},\hat{\mathbf{q}})$,
where $\mathbf{q}=[1, \cdots, 1, 0]\in{\mathbb{R}^{C+1}}$. 
% \vspace{-1ex}
\section{Experimental settings}
\label{sec-sxperimental-settings}
% \vspace{-1mm}
\subsection{Datasets}
\label{ssec-datasets}
In our research scenario, multiple speakers coexist in a mixed signal, each contributing multiple utterances. To synthesize data consistent with this scenario, we use speech from the LibriSpeech corpus~\cite{panayotov2015librispeech}, which provides the diarization labels for model training and evaluation. Noise signals are from the WHAM! dataset~\cite{Wichern2019WHAM}.
We consider four scenarios: (a) anechoic, (b) noisy, (c) reverberant, and (d) noisy and reverberant mixtures. Each scenario involves mixtures containing either two or three speakers, with a random number of utterances per speaker, resulting in eight dataset configurations. Each configuration comprises a training set, a validation set, and a test set, with 20,000, 2,000, and 2,000 mixtures, respectively. 

When synthesizing each anechoic mixture, we randomly select $C$ speakers from LibriSpeech. For each selected speaker, we randomly extract between $1$ and $5$ speech utterances, where the number of utterances follows a uniform distribution. We use the original Librispeech signal levels without scaling, where the relative level between any two speech signals varies within $[0, 5]$ dB.
A random silence interval is inserted in the beginning and between each utterance which duration is drawn uniformly from $[0, 3]$ seconds. Concatenating all speech utterances and silences for each speaker yields their complete speech signal $\mathbf{x}_c$. The anechoic multi-speaker, multi-utterance mixture is then obtained by superimposing these signals. 
This synthesis approach results in an average overlap ratio of $22.5\%$ in our 2-speaker dataset, which aligns with natural conversational scenarios.
% An example is shown in Fig.~\ref{fig-sample}(a).
To incorporate noise, we randomly extract a segment of equal length from the WHAM! noise dataset and adjust its amplitude according to a randomly generated signal-to-noise ratio (SNR) between $0$ and $10$ dB.
In the SNR calculation, the signal level is calculated as the average power of all speakers in the logarithmic scale.
This noise is then added to the anechoic mixture to produce a noisy multi-speaker, multi-utterance mixture.
To simulate reverberant mixtures, we randomly generate room sizes and reverberation time (RT$60$), along with random positions for the microphone and the $C$ speakers. Using the gpuRIR toolkit~\cite{diaz2021gpurir}, we apply reverberation to each speaker. Then we can generate a reverberant, multi-speaker, multi-utterance mixture.
The noise is added using the same method as before, resulting in a noisy, reverberant, multi-speaker, multi-utterance mixture.
All synthesized signals are single-channel with a sampling rate of 16 kHz. The training and test sets do not share speakers, noise signals, or rooms.

We modeled rooms with dimensions randomly selected from the following ranges: length $[4.0, 8.0]$ m, width $[4.0, 8.0]$ m, and height $[3.0, 4.0]$ m. The RT$60$ was varied between $0.2$ and $0.6$ seconds. Microphones were positioned at heights ranging from $1.0$ to $1.5$ m, while speakers were generated at heights between $1.5$ and $2.0$ m.
Microphones and speakers were restricted to be at least $0.5$ m away from any walls and each other.
\begin{table}[t]
  \caption{{$\Delta$SI-SDR (dB)}($\uparrow$) on the 2-speaker dataset.}
  \vspace{-2mm}
  \label{tab-SI-SDR-2spk}
  \centering
  \resizebox{0.47\textwidth}{!}{
  \begin{tabular}{ l| c c c c }
    \toprule
    \multicolumn{1}{c|}{\textbf{Models}} & 
    \multicolumn{1}{c}{\textbf{Anechoic}} & 
    \multicolumn{1}{c}{\textbf{Noisy}} & 
    \multicolumn{1}{c}{\textbf{Reverb}} & 
    \multicolumn{1}{c}{\textbf{Noisy \& Reverb}} \\
    % & & & & \textbf{Reverb} \\
    \midrule
    Conv-TasNet~\cite{luo2019conv}  & 6.1 & 6.0 & 3.5 & 4.0  \\
    Sepformer~\cite{Subakan2021} & 8.1 & 7.4 & 7.1  & 6.8  \\
    \midrule
    Recursive-SS~\cite{Takahashi2019}  & 7.1 & 6.4 & 4.5 & 3.5  \\
    EEND-SS~\cite{Maiti2023EENDSS}  & 7.7 & 7.2 & 5.7 & 5.5  \\
    SepEDA~\cite{Chetupalli2022} & 10.1 & 8.4 & 7.5  & 7.6  \\
    SepTDA~\cite{lee2024boosting} & 8.5 & 7.7 & 6.9 & 6.8  \\
    \midrule
    A-DCSS  & \textbf{11.2} & \textbf{9.7} & \textbf{8.9} & \textbf{8.7}  \\
    \bottomrule
  \end{tabular}
    }
    \vspace{-4mm}
\end{table}
\begin{table*}[t]
    \caption{Evaluation on 2-speaker \& 3-speaker dataset. DER (\%)($\downarrow$) is  diarization error rate. SCA(\%)($\uparrow$) is speaker counting accuracy.}
    \vspace{-0.2cm}
    \label{tab-flexible-spks}
    \centering
    \setlength{\tabcolsep}{5pt}
    \begin{tabular}{lccc|ccc|ccc|ccc}
    % \begin{tabular}{clccc|ccc|ccc|ccc}
    % \begin{tabular}{clcccc|cc|cccc|cc}
    \hline
    % \\
    \noalign{\vskip 2pt}
           & \multicolumn{3}{c}{\textbf{Anechoic}} &
         \multicolumn{3}{c}{\textbf{Noisy}} &  \multicolumn{3}{c}{\textbf{Reverb}} &  \multicolumn{3}{c}{\textbf{Noisy \& Reverb}} \\
        \cmidrule{2-13}
          \textbf{Models} & $\Delta$SI-SDR & DER  &  SCA  & $\Delta$SI-SDR & DER & SCA & $\Delta$SI-SDR & DER  &  SCA  & $\Delta$SI-SDR & DER & SCA   \\ 
        \toprule
          Recursive-SS~\cite{Takahashi2019} & 6.2 & -- & -- & 5.5 & -- & -- & 3.9 & -- & -- & 3.1 & -- & -- \\
                                  EEND-SS~\cite{Maiti2023EENDSS} &  6.7  & 8.4 &  96.7  & 6.0 & 9.7 &  94.1  & 4.6 &  8.5 & 96.2  & 4.5 & 10.2 & 93.7 \\
                                 SepEDA~\cite{Chetupalli2022} &  8.6  & -- & 97.5  & 7.3 &  -- & 96.3 & 6.4 & -- & 96.9 & 6.7 & -- & 96.1 \\ 
                                  SepTDA~\cite{lee2024boosting} &  7.8  & -- & \textbf{98.7} & 6.9 & -- & \textbf{97.8} & 6.0 & --  & \textbf{98.2} & 5.9 & -- & \textbf{97.5}  \\ 
                                  A-DCSS &  \textbf{9.7}  & \textbf{5.9} & 97.9  & \textbf{8.8} & \textbf{6.4} & 96.7 & \textbf{7.8} & \textbf{6.2}  & 97.1 & \textbf{7.7} & \textbf{7.1} & 96.4 \\ 
                            
    \bottomrule
    \end{tabular}
    \vspace{-1.5em}
\end{table*}
\begin{table}[t]
  \caption{Ablation experiment.}
  \vspace{-2mm}
  \label{tab-ablation}
  \centering
  \resizebox{0.47\textwidth}{!}{
  \begin{tabular}{ l| c c c c }
    \toprule
    \multicolumn{1}{c|}{\#} & 
    \multicolumn{1}{c}{\textbf{Attractor}} & 
    \multicolumn{1}{c}{\textbf{Diarization}} & 
    % \multicolumn{1}{c}{\textbf{Params ($M$)}} & 
    \multicolumn{1}{c}{\textbf{Counting}} &
    \multicolumn{1}{c}{$\Delta$\textbf{SI-SDR}} \\
    % & & & & \textbf{Reverb} \\
    \midrule
    1  & -- & --  & -- & 8.2 \\
    2 & \text{Transformer} & --   & \checkmark & 8.8 \\
    3  & \text{Transformer} & \checkmark   & \checkmark & 10.3 \\
    4  & \text{RNN} & -- &  \checkmark & 10.1  \\
    5 & \text{RNN} & \checkmark  & \checkmark & \textbf{11.2}  \\
    \bottomrule
  \end{tabular}
    }
    \vspace{-6mm}
\end{table}
\vspace{-1ex}
% \vspace{-1mm}
\subsection{Training Configurations}
\label{ssec-training-configurations}
The A-DCSS model was trained in two phases: first with two-speaker samples, and then with a dataset containing a varying number of speakers.
We utilized the ESPNet toolkit~\cite{watanabe2018espnet} for model training, employing the Adam optimizer.
For the first phase, the initial learning rate was set to $1.0 \times 10^{-3}$, while in the second phase, it was reduced to $1.0 \times 10^{-5}$. Both training phases were configured to run for a maximum of 200 epochs with an early stopping strategy.
The model architecture was configured with an encoder/decoder kernel size $L$ of $16$ and a stride of $8$ samples. The feature dimensions $D$ and $F$ were both set to $256$. The chunk length $K$ is $96$. We used $4$ attention heads in each transformer layer. The $N_{\text{triple}}$ was set to $6$. The thresholds $\tau_{\text{exist}}$ and $\tau_{\text{diar}}$ were both set to $0.5$. The $\lambda_{\text{s}}$, $\lambda_{\text{d}}$, and $\lambda_{\text{e}}$ were set to $0.8$, $0.1$, and $0.1$, respectively.
All these hyperparameters were determined through experiments on the validation set.
During training, the batch size was set to $4$, each training sample was split into $10$ second segments in a batch.
\vspace{-1ex}
% \vspace{-1mm}
\subsection{Evaluation Metrics}
\label{ssec-dvaluation-metrics}
We evaluate the separation performance of the A-DCSS using the scale-invariant source-to-distortion ratio improvement ($\Delta$SI-SDR)~\cite{LeRoux2018a}. Diarization performance is measured by the diarization error rate (DER)~\cite{fiscus2006rich}, which assesses the accuracy of detecting speaker activity. We also utilize speaker counting accuracy (SCA)~\cite{chetupalli2023speaker} to evaluate the counting performance.
% \vspace{-1ex}
\section{Experimental Results}
\label{sec-Experimental-Results}
We evaluated the performance of the A-DCSS in two scenarios: one with a fixed number of speakers and the other with a varying number of speakers. Besides, an ablation study was conducted to analyze the importance of architectural choices.
We compare the representative single-channel source separation methods, including Conv-Tasnet~\cite{luo2019conv} and Sepformer~\cite{Subakan2021}.
In addition, we evaluate four advanced models designed for speech separation in scenarios where the number of speakers is unknown, as shown in Table \ref{tab-SI-SDR-2spk}.
\begin{itemize}
    \item Recursive-SS~\cite{Takahashi2019}: Employs a recursive structure. The separation component is based on Conv-Tasnet~\cite{luo2019conv}.
    \item EEND-SS~\cite{Maiti2023EENDSS}: A baseline system integrating speaker diarization, counting, and separation. Both EEND-SS and A-DCSS adopt a three-task system with similar RNN attractors. The difference is that EEND-SS uses TCN~\cite{luo2019conv} for feature extraction and separation, while A-DCSS employs transformer-based feature embedding module and separator, along with FiLM~\cite{Perez2018film} for feature fusion.
    \item SepEDA~\cite{Chetupalli2022}: Integrates speaker counting and separation. Uses an RNN attractor and a separator based on Sepformer~\cite{Subakan2021}. In SepEDA~\cite{Chetupalli2022}, the attractor is used only for speaker counting, whereas in A-DCSS, it is utilized for both speaker counting and activity detection. Furthermore, the design of the separation component and the fusion of the attractor features differs between SepEDA~\cite{Chetupalli2022} and A-DCSS.
    \item SepTDA~\cite{lee2024boosting}: Combines speaker counting and separation using a transformer attractor and transformer-based separator. Like A-DCSS, its separation module is inspired by~\cite{Subakan2021}. The difference is that SepTDA employs transformer attractors for joint counting and separation, while A-DCSS uses RNN attractors for diarization, counting, and separation.
\end{itemize}
\vspace{-1ex}
\subsection{Fixed Number of Speakers}
\label{ssec-exp-fixed-spk}
We conducted experiments using synthesized datasets of two-speaker mixtures in four acoustic scenarios, as shown in Table~\ref{tab-SI-SDR-2spk}.
In the anechoic condition, the A-DCSS achieves the best performance. Comparing Conv-TasNet~\cite{luo2019conv}, Recursive-SS~\cite{Takahashi2019}, and EEND-SS~\cite{Maiti2023EENDSS}, we observe that incorporating recursive structures (Recursive-SS~\cite{Takahashi2019}) or attractor modules (EEND-SS~\cite{Maiti2023EENDSS}) enhances separation performance in multi-speaker multi-utterance scenarios.
The attractor outperforms the recursive process. In noisy, reverberant, and noisy-reverberant environments, we observe similar trends. Both noise and reverberation cause performance degradation of the A-DCSS, and the impact of reverberation is greater than that of noise. 
\vspace{-1ex}
% \vspace{-1mm}
\subsection{Unknown Number of Speakers}
\label{ssec-exp-flexible-spk}
We evaluated the A-DCSS on a varying number of speakers dataset combining 2-speaker and 3-speaker scenarios. Due to potential speaker counting errors, the number of reference signals may differ from the number of separated speech signals. To address this, we added silent audio signals to either the reference or separated speech signals to match the signal count. Table \ref{tab-flexible-spks} presents the results.
In all conditions tested, our proposed method achieved the best performance. All models showed performance degradation compared to the scenario with the fixed number of speakers. For the DER evaluation, the A-DCSS significantly outperformed the EEND-SS~\cite{Maiti2023EENDSS} baseline. Regarding the accuracy of source counting, the SepTDA method~\cite{lee2024boosting} performed best, with our proposed method ranking second. This comparison suggests that the transformer attractor outperforms the RNN attractor for the source count estimation subtask. For speaker diarization and counting, reverberation had little impact, while noise had a clearer effect.
\vspace{-1ex}
% \vspace{-1mm}
\subsection{Ablation experiment}
\label{ssec-exp-ablation}
In multi-utterance scenarios, the attractor module is essential to the A-DCSS. We conducted ablation experiments using the two-speaker anechoic dataset, with results in Table \ref{tab-ablation}. The transformer attractor follows the SepTDA~\cite{lee2024boosting}, and the RNN attractor is detailed in this paper. Diarization and counting represent two possible output branches of the attractor. All other modules and parameters remained unchanged during ablation.

When comparing row $1$ to the others, the effect of the attractor is significant. 
Comparing row $2$ and row $3$, it can be observed that the $\Delta$SI-SDR improves by $1.5$ dB when additional diarization constraints are added to the transformer attractors. Similarly, comparing row $4$ and row $5$ (A-DCSS), the SI-SDR increases by $1.1$ dB when diarization output is incorporated into the RNN attractors. This demonstrates the importance of speaker diarization outputs from attractors for separation in multiple utterance scenarios.
Further analysis of the results between row $2$ and row $4$, as well as between row $3$ and row $5$, indicates that the RNN-based attractor outperforms the transformer-based attractor in our experimental setup. This observation is also supported by the results presented in Table \ref{tab-SI-SDR-2spk}.
\vspace{-1ex}
% \vspace{-1mm}
\section{Conclusions}
\label{sec-conclusions}
In this paper, we proposed a novel single-channel joint speech separation system that effectively handles scenarios with an unknown number of speakers, where each speaker may contribute multiple utterances. The system employs an RNN attractor module to estimate the number of sources and detect source activity, resulting in significant improvements in separation performance. Additionally, the proposed method demonstrates excellent robustness in noisy and reverberant environments. 

\newpage
\bibliographystyle{IEEEtran}
\bibliography{Bib_YWang}

\end{document}